\documentclass[11pt]{article}
\usepackage{epsf}

\begin{document}

\title{Computation of a homogeneous coordinate time series for European
 GPS stations by reprocessing of the weekly EPN solutions}
\author{Natalia Panafidina, Zinovy Malkin \\
 \it Institute of Applied Astronomy RAS, nab. Kutuzova 10,
 St.Petersburg 191187, Russia}
\date{May 28, 2003}
\maketitle

\begin{abstract}

Weekly coordinate time series for the all EPN (European Permanent GPS Network)
stations for the period of observations beginning from GPS week 834 was
obtained by a reprocessing of the official EPN solutions.
Comparison of this series with EPN Central Bureau (EPN CB)
and the previous IAA two-year solution for selected EPN stations
obtained by reprocessing of the original observations is performed.
Results of comparison show that new solution based on fiducial-free
strategy is most likely free of seasonal errors.
Now the new IAA EPN solution is being computed regularly
(on availability of the official EPN solution)
and is available to any interested group.
\end{abstract}

\section{Introduction}

GPS observations collected from the European GPS network are widely used for
geodesy and geodynamic researches in the region.
The network is coordinated by the EPN CB which also provides the official
analysis of the observations.
Two EPN weekly solutions are available.
The first one is computed at the Bundesamt f\"ur Kartographie und
Geod\"asie, Germany (previously at the Center for Orbit Determination in Europe,
Astronomical Institute of the University of Bern, Switzerland) and distributed
as SINEX files.  Hereafter this solution is referred to as EUR.
Unfortunately, this solution is not suitable for high-accuracy geodesy
applications because the using of fiducial approach and periodic change of
the reference system cause jumps in coordinates and distortion of the network.
Direct use of this solution can lead to some confusions (see e.g.
\cite{Lanotte99}).
Besides the EUR solution does not contain any information about
displacement of the fiducial stations.

The second solution is computed at the EPN CB by reprocessing the
previous one and seems to provide high-quality information about
the movement of the all EPN stations.  Unfortunately, this solution
is not distributed in SINEX files, and it is
difficult to use it in scientific analysis.

For these reasons, several years ago the IAA undertook a special project
aimed at the computation of an independent coordinate time series
for all the EPN stations.
A detailed description of the project was given in \cite{Malkin01o}.

At the first stage the original GPS observations for a selected EPN
subnetwork at two-year interval were reprocessed using fiducial-free strategy.
Basic theoretical background for this approach can be found in
\cite{Blewitt92,Heflin92,Springer95,Zumberge97,Dong98,MalkinVoinov01,Malkin01n}.
Previously this strategy was tested during processing of two Baltic
Sea Level campaigns \cite{Springer95,VoinovMalkin99}
Obtained coordinate time series appeared to be more stable in sense of random
and systematic errors \cite{MalkinVoinov01,Malkin01n}.
It is hereafter referred to as I1.

Unfortunately, this way of reprocessing requires too much resources and we
tried two another approaches.
The first one is based on reprocessing of existing EUR solutions
(the second stage of the IAA project).
The processing strategy is described below.
Obtained coordinate time
series was compared with other EPN solutions and the previous IAA
solution I1. Results of comparison show that
obtained solution is most likely free of seasonal errors.

At the third stage of the our project we plan to obtain an EPN coordinate
time series by an independent combination of individual solutions
provided by the EPN Analysis Centers, also using fiducial-free approach.
This work is under development and is planned to be completed by the
end of 2003.

\section{Processing strategy}

Our processing was made in two steps.
At the first one the EUR solutions are de-constrained using
the a priori coordinates and covariance matrices contained in the SINEX files
following the strategy proposed in \cite{Brockmann96}.

After this the transformation of the obtained free network solution to the
ITRF2000 is made.  Unlike EPN CB strategy, we use for the transformation
all the stations presented in the solution using weights
dependent on their position accuracy (taking into account both
errors in position and velocity) in the EUR solution and
in the reference ITRF catalogue.

Two solutions were computed:
with 6 and 7-parameter Helmert transformation.
Solution obtained with 6-parameter
transformation is hereafter referred to as I2, solution obtained with
7-parameter transformation is referred to as I3.

\begin{figure}
\centerline{\epsfclipon \epsfxsize=130mm \epsfbox{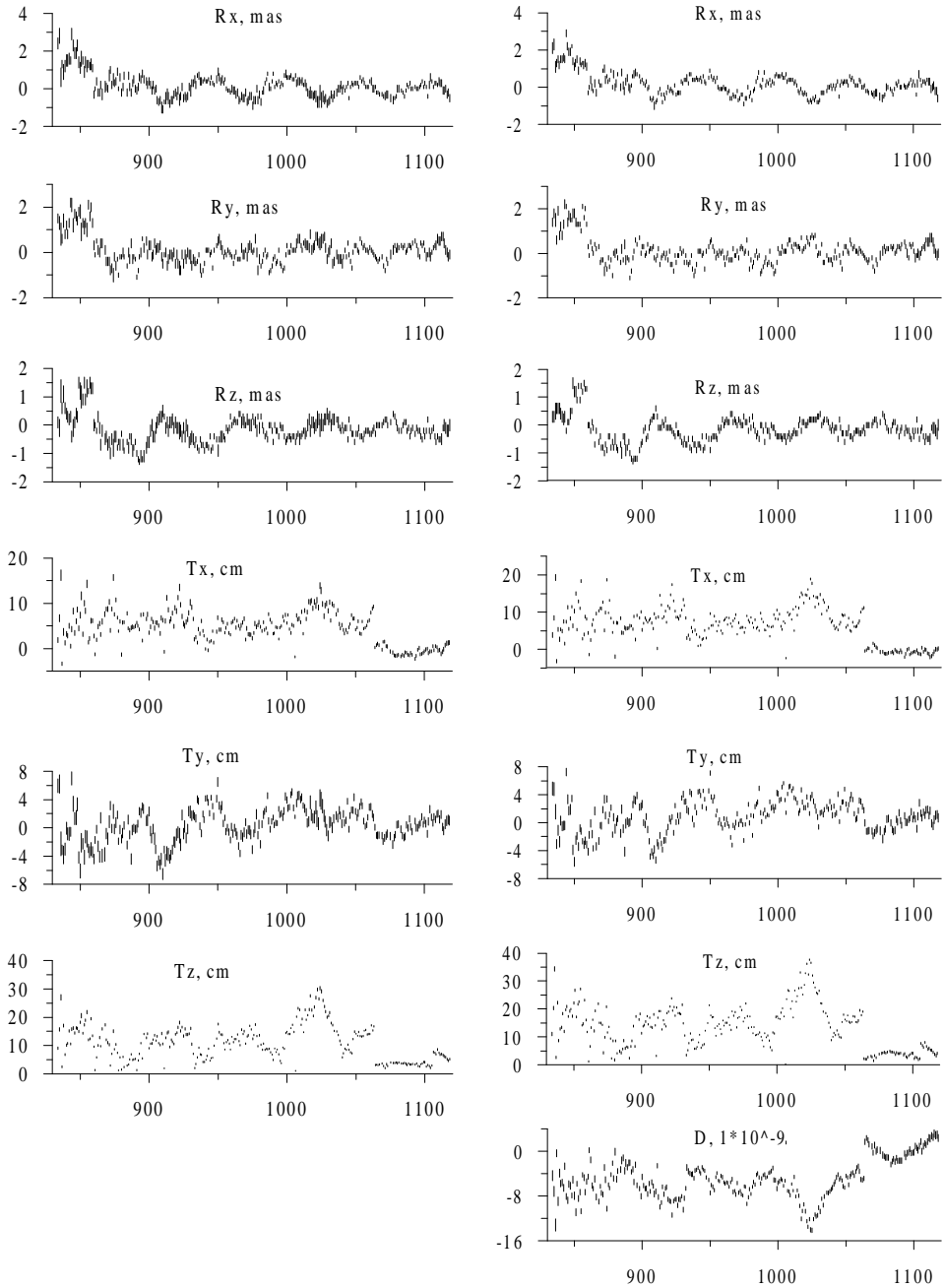}}
\caption{Helmert parameters time series for transformation w.r.t. ITRF2000:
with 6 parameters (left) and with 7 parameters (right) (R~--- rotation parameters,
T~--- translation parameters, D~--- scale); vertical bars on the plots present parameters
errors.}
\label{fig:tr_helm}
\end{figure}

Time series of
transformation parameters are shown in Figure~\ref{fig:tr_helm}.
One can clearly see the seasonal and other peculiarities in these time
series which should be investigated separately.

Analysis of the solutions obtained by this method showed that
the obtained coordinate time series have a similar quality with
the EPN CB solution with some small discrepancies which can be
explained by details of used approaches.
However analysis of the errors in the station coordinates
reveals very large irregularities, caused most probably by
inconsistency in SINEX blocks (e.g. wrong scaling of covariance
matrices).

\begin{figure}[ht]
\centerline{\epsfclipon \epsfxsize=100mm \epsfbox{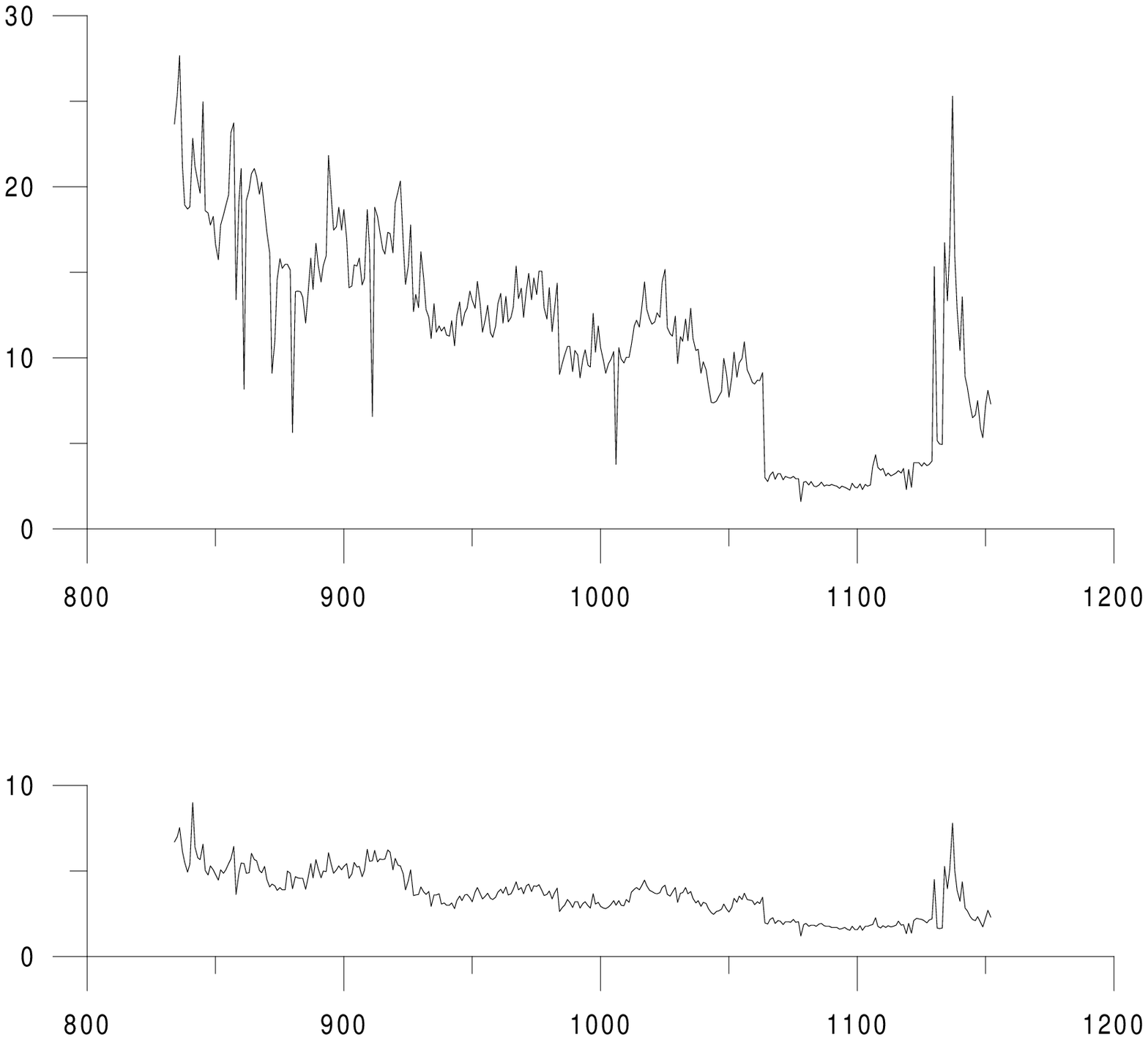}}
\caption{Mean coordinate errors without (top) and with (bottom) re-scaling, mm;}
\label{fig:errors}
\end{figure}

To reduce the inconsistency of coordinate errors we used
a re-scaling of computed covariance matrices.
The re-scaling factor was chosen in such a way that the mean coordinate error
of non-fiducial stations in the EUR solution is equal to the mean coordinate
error of the same stations of the new solution.
The plots of mean coordinate errors of our solution with and without re-scaling
of covariance matrices are shown in Figure~\ref{fig:errors}.

\section{Comparison and conclusions}

\begin{figure}
\fboxrule 2pt
\centerline{\fbox{\epsfclipon \epsfxsize=75mm \epsfbox{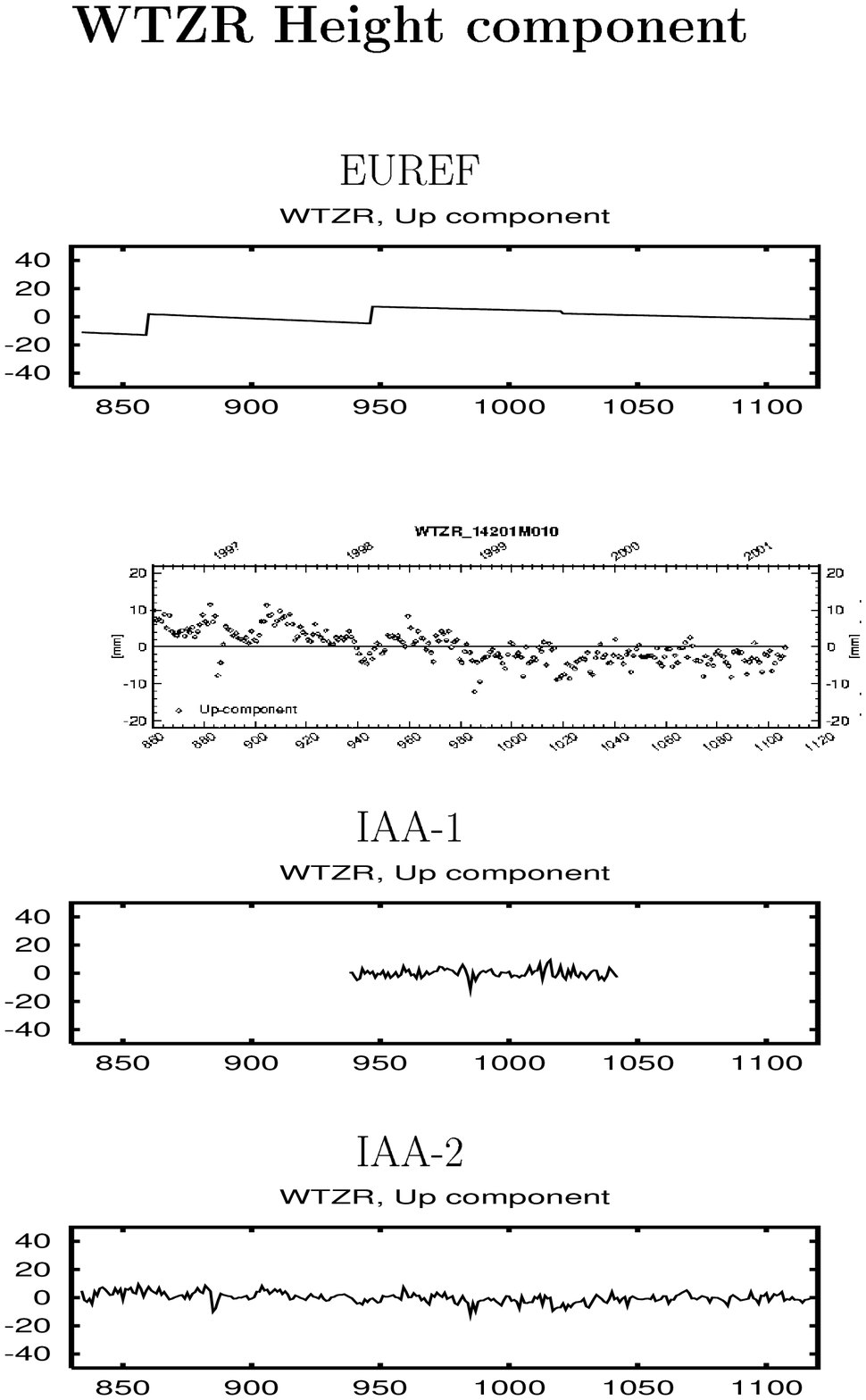}}
 \hskip 2mm \fbox{\epsfclipon \epsfxsize=75mm \epsfbox{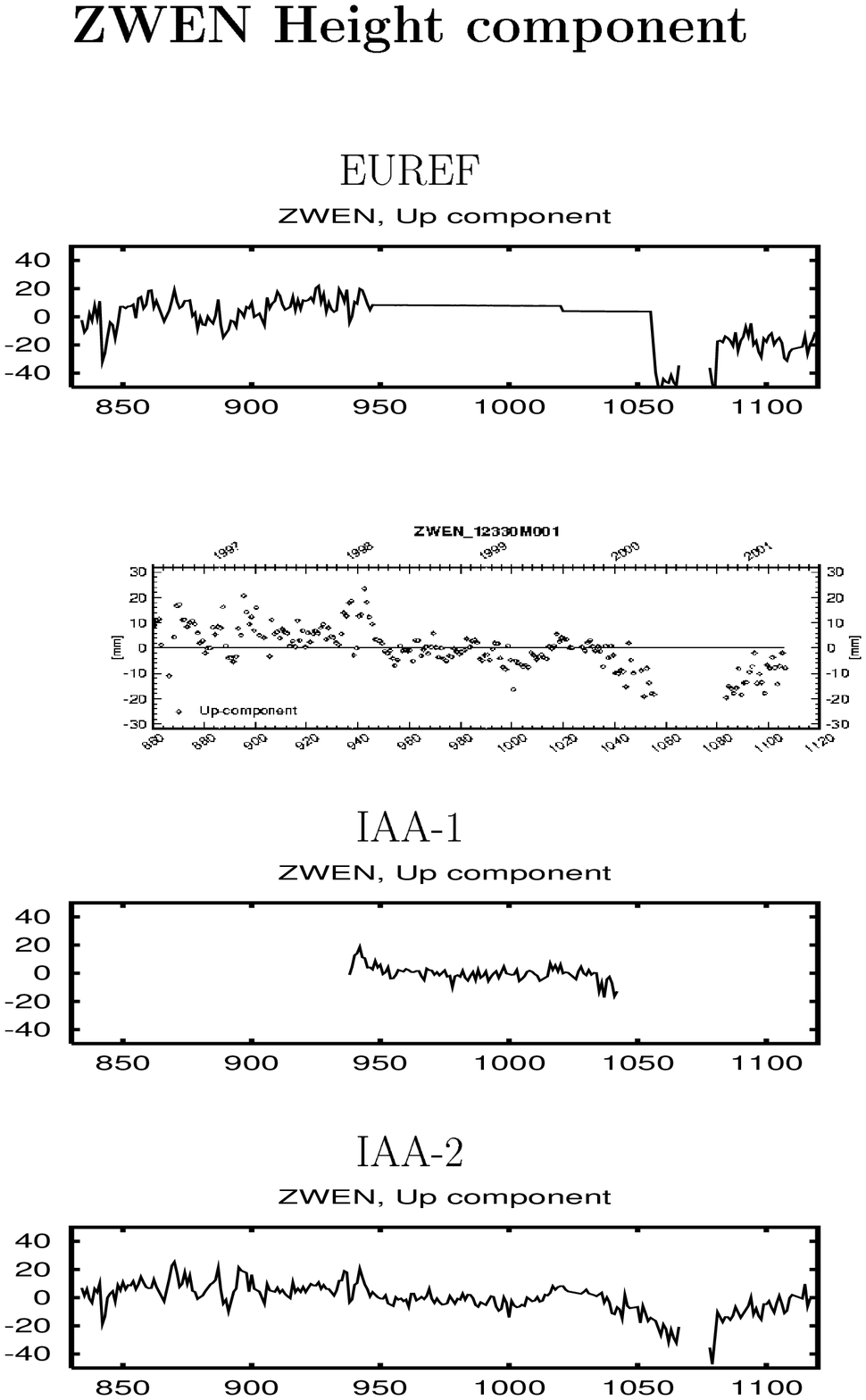}}}
\vskip 2mm
\centerline{\fbox{\epsfclipon \epsfxsize=75mm \epsfbox{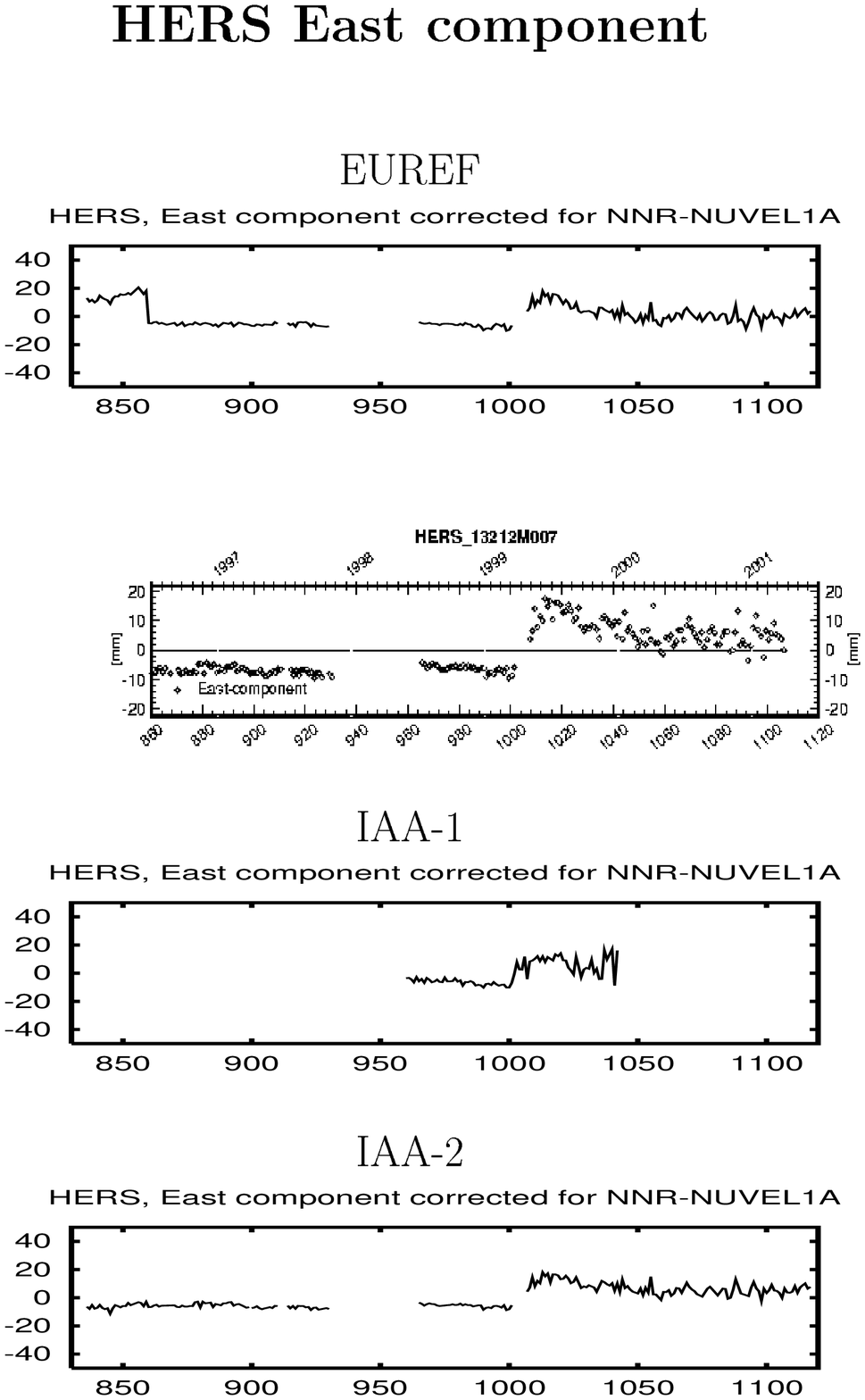}}
 \hskip 2mm \fbox{\epsfclipon \epsfxsize=75mm \epsfbox{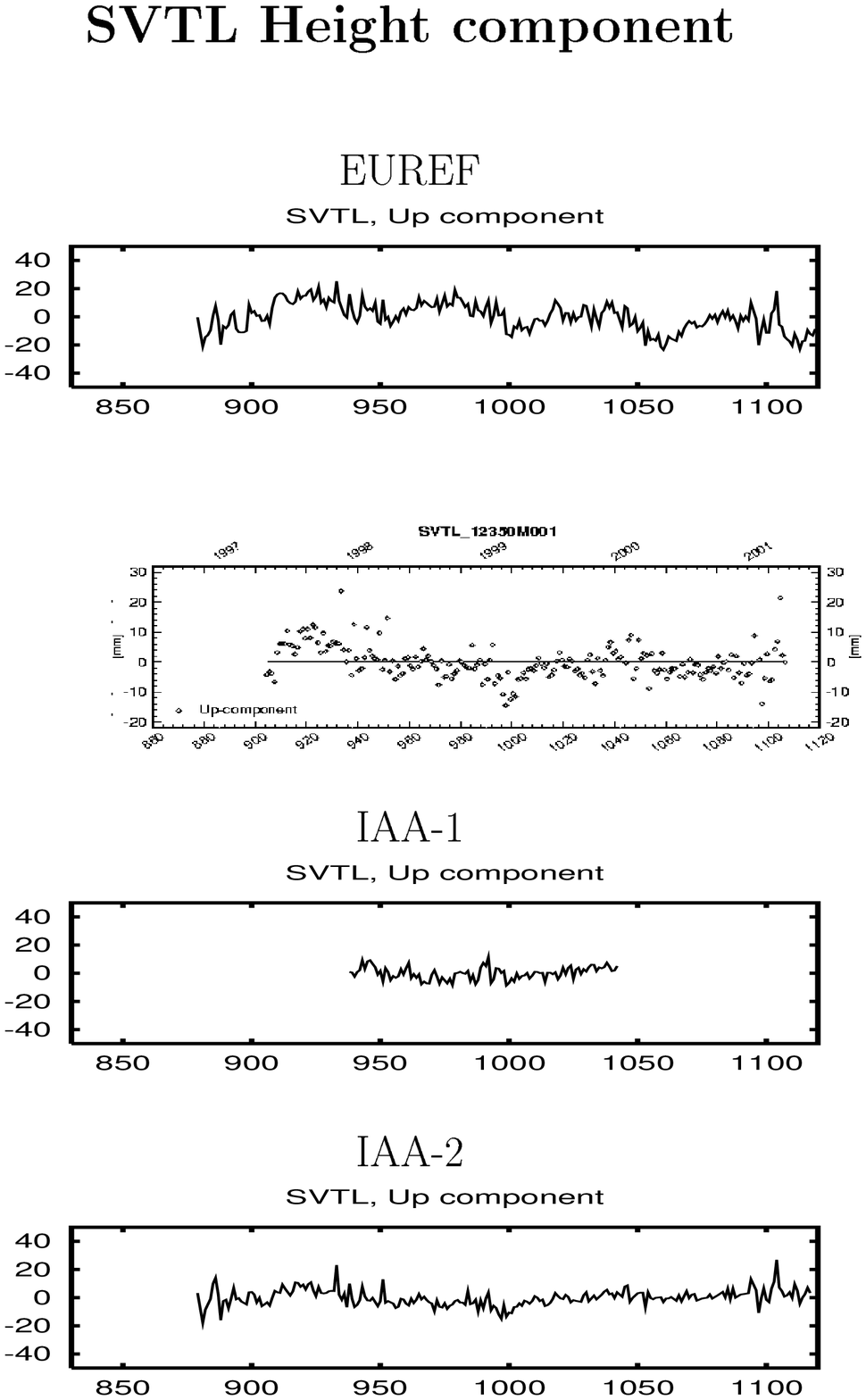}}}
\caption{Some examples of the coordinate time series, mm
(top to down: EUR solution, EUREF CB solution (copied from the web page
\cite{EPN}), I1 solution, and I2 solution.).}
\label{fig:series}
\end{figure}

Six-year coordinate time series for all EPN stations were
computed and compared with two other solutions (EUR solution and our previous
two-year solution for selected european stations). Several examples are
presented in Figure~\ref{fig:series}.
It is seen that all series provide determination of main details in
behavior of station position.

A comparison of four solutions mentioned above is presented in
Table~\ref{tab:1}. The table contains results of determination of
week-to-week repeatability (Allan variance) interpreted as random error and
amplitude of seasonal term in variation of station coordinates in the local
ENU system.

\begin{table}
\centering
\caption{Statistics for 10 stations presented in all solutions.}
\label{tab:1}
\bigskip
\tabcolsep=12pt
\begin{tabular}{|c|c|c|c|c|c||c|c|c|c|}
\hline
\multicolumn{2}{|c|}{Station}&\multicolumn{4}{|c|}{Allan variance, mm}&
 \multicolumn{4}{|c|}{Annual term, mm} \\
\cline{3-10}
\multicolumn{2}{|c|}{ }& E  & I1  & I2  & I3  &  E  & I1  & I2  & I3 \\
\hline
GLSV  & dE & 1.2 & 1.2 & 2.0 & 1.6 & 1.0 & 0.6 & 5.7 & 1.2 \\
      & dN & 1.1 & 1.1 & 1.1 & 1.3 & 1.7 & 1.4 & 1.7 & 1.3 \\
      & dH & 3.7 & 3.0 & 2.4 & 2.0 & 7.0 & 3.5 & 3.5 & 2.7 \\
JOZE  & dE & 0.9 & 0.9 & 1.0 & 1.8 & 0.7 & 0.9 & 1.5 & 0.5 \\
      & dN & 0.8 & 0.8 & 0.8 & 0.8 & 0.9 & 0.4 & 0.3 & 0.5 \\
      & dH & 3.1 & 2.9 & 2.5 & 1.9 & 3.7 & 1.6 & 0.9 & 1.3 \\
LAMA  & dE & 1.2 & 1.1 & 1.1 & 1.5 & 0.4 & 0.4 & 2.3 & 0.4 \\
      & dN & 0.9 & 0.9 & 0.9 & 1.0 & 0.3 & 0.7 & 1.1 & 0.7 \\
      & dH & 2.9 & 3.0 & 2.0 & 1.6 & 5.6 & 2.8 & 2.5 & 1.9 \\
MDVO  & dE & 1.5 & 1.4 & 2.2 & 2.5 & 0.3 & 0.6 & 5.2 & 0.2 \\
      & dN & 1.1 & 1.3 & 1.1 & 2.3 & 0.5 & 1.0 & 0.8 & 1.0 \\
      & dH & 5.6 & 5.1 & 5.4 & 4.3 & 6.6 & 6.4 & 1.2 & 1.3 \\
MEDI  & dE & 1.8 & 1.7 & 2.0 & 3.2 & 1.9 & 1.8 & 1.6 & 1.8 \\
      & dN & 2.4 & 2.6 & 2.5 & 2.0 & 1.9 & 2.2 & 1.4 & 3.4 \\
      & dH & 2.7 & 2.8 & 2.7 & 1.8 & 3.3 & 1.7 & 2.0 & 1.3 \\
METS  & dE & 1.1 & 1.3 & 1.4 & 1.8 & 2.2 & 0.3 & 2.8 & 0.5 \\
      & dN & 1.8 & 1.9 & 2.0 & 1.2 & 2.6 & 1.6 & 2.9 & 1.1 \\
      & dH & 4.0 & 3.5 & 2.9 & 2.8 & 4.9 & 0.9 & 3.5 & 3.0 \\
NOTO  & dE & 1.2 & 1.4 & 1.1 & 2.1 & 1.1 & 0.8 & 1.4 & 0.9 \\
      & dN & 1.3 & 1.5 & 2.2 & 1.0 & 3.0 & 0.9 & 7.5 & 1.7 \\
      & dH & 3.0 & 4.7 & 2.7 & 2.0 & 2.2 & 4.1 & 2.3 & 3.1 \\
SVTL  & dE & 1.3 & 1.7 & 1.8 & 1.9 & 1.1 & 1.1 & 4.2 & 0.7 \\
      & dN & 1.0 & 1.2 & 1.3 & 1.2 & 1.0 & 0.6 & 2.7 & 0.5 \\
      & dH & 4.4 & 3.2 & 3.4 & 3.0 & 6.9 & 1.9 & 2.6 & 1.9 \\
WSRT  & dE & 0.6 & 0.9 & 0.8 & 1.3 & 0.4 & 0.6 & 0.9 & 0.1 \\
      & dN & 0.8 & 1.1 & 0.9 & 0.7 & 0.3 & 0.6 & 0.9 & 0.6 \\
      & dH & 2.3 & 2.7 & 1.9 & 1.6 & 1.0 & 2.1 & 2.0 & 2.3 \\
ZECK  & dE & 1.4 & 1.5 & 2.6 & 2.1 & 2.1 & 0.6 & 7.8 & 2.4 \\
      & dN & 1.4 & 1.7 & 2.0 & 1.7 & 1.2 & 1.0 & 5.0 & 1.9 \\
      & dH & 3.8 & 2.6 & 3.8 & 2.8 & 11.8& 0.5 & 5.3 & 3.2 \\
\hline
Mean  & dE & 1.3 & 1.6 & 1.6 & 1.1 & 1.6 & 1.2 & 3.3 & 0.9 \\
      & dN & 1.2 & 1.5 & 1.5 & 1.2 & 1.4 & 1.1 & 2.4 & 1.3 \\
      & dH & 3.5 & 3.5 & 3.0 & 2.9 & 5.0 & 2.6 & 2.6 & 2.2 \\
\hline
\end{tabular}
\end{table}

Comparison shows that I3 solution obtained using 7-parameter Helmert
transformation to the ITRF2000 provides minimum random
error and seasonal variations. Of course, the latter may mean merely loss of
geophysical signal, but comparison with EPN CB series (available at the EPN
Web site) and global solutions (T.~Springer, private communication)
shows that most likely seasonal terms
observed in the EPN solution is caused by systematic errors induced by errors
in modeling of position of fiducial stations.

Evidently, more thorough
consideration should be made to make a choice between using 6 or 7-parameter
Helmert transformation of free network solution to ITRF. The first impression
is that it is more reasonable to apply 6-parameter transformation to a global
network, whereas 7-parameter transformation is more adequate to regional data.

Now the IAA EPN solutions are being computed regularly
using described strategy,
on availability of the official EPN solutions,
and the results are available to any interested group
(on request at the moment, later it will be put to the Internet).

\section{Aknowledgments}
Authors are very grateful to Matthias Becker and Daniel
Ineichen for their help in understanding of strategy of
computation of the EPN combined solutions.

\end{document}